\documentclass[12pt]{article}
\usepackage{epsf}
\usepackage{amsfonts} 
\usepackage{amsmath} 
\usepackage[latin2]{inputenc} 
\usepackage[T1]{fontenc} 
\usepackage{epsfig} 
 
\newlength{\dinwidth}                          
\newlength{\dinmargin} 
\def\cs{{\cal S}}
\def\co{{\cal O}}                                                               
           
\def\beq{\begin{equation}}  
\def\eeq{\end{equation}}  
\def\bea{\begin{eqnarray}}   
\def\eea{\end{eqnarray}}   
\def\bq{\begin{quote}}   
\def\eq{\end{quote}}   
\def\bi{\begin{itemize}}   
\def\ei{\end{itemize}}   
\def\beqa{\begin{eqnarray}}   
\def\eeqa{\end{eqnarray}}   
\def\be{\begin{enumerate}}   
\def\ee{\end{enumerate}}   
\def\bi{\begin{itemize}} 
\def\ei{\end{itemize}}

\def\pa{\partial}

\def\r2{\sqrt{2}}   
 
\def\bi{\begin{itemize}}   
\def\ei{\end{itemize}}

\def\beq{\begin{equation}}  
\def\eeq{\end{equation}}  
\def\bea{\begin{eqnarray}}   
\def\eea{\end{eqnarray}}   
\def\bq{\begin{quote}}   
\def\eq{\end{quote}}   
\def\bi{\begin{itemize}}   
\def\ei{\end{itemize}}   
\def\beqa{\begin{eqnarray}}   
\def\eeqa{\end{eqnarray}}   
\def\be{\begin{enumerate}}   
\def\ee{\end{enumerate}}   
\def\beq{\begin{equation}}   
\def\eeq{\end{equation}}   
\def\bi{\begin{itemize}} 
\def\ei{\end{itemize}}  
\def\bc{\begin{center}} 
\def\ec{\end{center}}

\def\pa{\partial}

\def\r2{\sqrt{2}}   
 
\def\bi{\begin{itemize}}   
\def\ei{\end{itemize}}

\def\beq{\begin{equation}}  
\def\eeq{\end{equation}}  
\def\bea{\begin{eqnarray}}   
\def\eea{\end{eqnarray}}   
\def\bq{\begin{quote}}   
\def\eq{\end{quote}}   
\def\bi{\begin{itemize}}   
\def\ei{\end{itemize}}   
\def\beqa{\begin{eqnarray}}   
\def\eeqa{\end{eqnarray}}   
\def\be{\begin{enumerate}}   
\def\ee{\end{enumerate}}   
\def\beq{\begin{equation}}   
\def\eeq{\end{equation}}   
\def\bi{\begin{itemize}} 
\def\ei{\end{itemize}}

\def\pa{\partial}

\def\r2{\sqrt{2}}   
  
\def\bi{\begin{itemize}}   
\def\ei{\end{itemize}}

\def\bc{\begin{center}} 
\def\ec{\end{center}}   
 
\DeclareMathAlphabet{\scr}{U}{rsfs}{m}{n}  
\begin{document} 
\pagestyle{empty}  
\begin{flushright} 
AEI-2002-059\\IFT-2002-35 
\end{flushright}  
\vskip 2cm  
\begin{center}  
{\Huge  
On Scherk-Schwarz mechanism\\
in gauged five-dimensional supergravity\\
and on its relation to bigravity} 
\vspace*{5mm} \vspace*{1cm}   
\end{center}  
\vspace*{5mm} \noindent  
\vskip 0.5cm 
\centerline{\bf Zygmunt Lalak${}^{1,2}$ 
and Rados\l aw Matyszkiewicz${}^1$}  
\vskip 1cm   
\centerline{\em ${}^{1}$Institute of Theoretical Physics}  
\centerline{\em University of Warsaw, Poland}  
\vskip 0.3cm
\centerline{\em ${}^{2}$Max-Planck-Institut f\"{u}r Gravitationsphysik}
\centerline{\em Golm, Germany}
\vskip 2cm  
%\centerline{\Huge DRAFT \today}  
\vskip1cm
\centerline{\bf Abstract}  
\vskip 0.5cm
We demonstrate the relation between the Scherk-Schwarz mechanism and flipped 
gauged brane-bulk supergravities in five dimensions. 
We discuss the form of supersymmetry violating Scherk-Schwarz terms
in pure supergravity and in supergravity coupled to matter. 
We point out that brane-induced supersymmetry breakdown in 5d Horava-Witten 
model is not of the Scherk-Schwarz type.
We discuss in detail flipped super-bigravity, which is the locally 
supersymmetric extension of the $(++)$ bigravity.     
\vskip3cm  
%\begin{flushleft}September 2002\end{flushleft}
\newpage  
\pagestyle{plain}   
\section{Introduction}%\input{gluon.tex}  
The issue of hierarchical supersymmetry breakdown in supersymmetric 
brane worlds is one of the central issues in the quest for 
a phenomenologically viable extra-dimensional extension of the Standard Model.
Many attempts towards formulating scenarios of supersymmetry breakdown 
that use new features offered by extra-dimensional setup have been made 
\cite{Horava:1996ma},\cite{Horava:1996vs},\cite{Antoniadis:1997xk},\cite{Antoniadis:1997ic},\cite{Dudas:1997jn},\cite{Nilles:1998sx},\cite{Lalak:1997zu},\cite{Lukas:1998tt},\cite{Ellis:1998dh},\cite{Falkowski:2001sq},\cite{Barbieri:2000vh},\cite{Bagger:2001ep},\cite{vonGersdorff:2002tj}.%\cite{Horava:1996ma}-\cite{vonGersdorff:2002tj}.
 One of them is supersymmetry breakdown triggered by imposing 
nontrivial boundary conditions on field configurations along the compact 
transverse dimensions usually referred to as the Scherk-Schwarz mechanism. 
In the work published so far the investigation concentrated on 
flat Minkowski-type geometries 
of the branes and bulk, neglecting the backreaction of the gauge sectors 
on the gravitational background (in particular assuming a fixed radii of the 
extra dimensions), and in general the interactions of 
various fields with (super)gravity (but see \cite{Alonso-Alberca:2000ne},
\cite{Falkowski:2001sq}). 
On the other hand, it is precisely partial 
`unification' of the Standard Model with gravity that makes the Brane World 
scenarios so intriguing and appealing. In this note we would like to clarify 
the status of the Scherk-Schwarz approach to supersymmetry breaking in the 
nontrivial gravity backgrounds using the class of simple warped gauged 
supergravities with flipped boundary conditions described in \cite{Brax:2001xf}. 

In particular, we find out that the simple flipped supergravity forms 
the locally supersymmetric extension of the $(++)$ bigravity model of 
Kogan et. al. \cite{Kogan:1999wc},\cite{Kogan:2000vb},\cite{Kogan:2000xc}. 
In such a setup one circumvents 
the van Dam-Veltman-Zakharov observation about the nondecoupling of 
the additional polarization states of the massive graviton  \cite{Kogan:2000uy},\cite{Porrati:2000cp},\cite{Karch:2001jb}. 
The size of the residual four-dimensional cosmological constant can be tuned 
to arbitralily small values by taking the distance between branes suitably 
large. We discuss the mass spectrum of gravitons and gravitini 
in the resulting super-bigravity.  
 
To begin with let us notice that the brane-world version of 
the Scherk-Schwarz mechanism contains a number of ingredients. 
Assuming for simplicity that the extra dimension is just a 
circle $S^1$, 
and that the higher-dimensional theory has a group G of global symmetries,
one can impose on various fields periodicity with a twist, that is instead of 
standard periodicity conditions $\psi(y+ 2 \pi \rho)=\psi(y)$ 
it is consistent to demand that after following the circle a configuration 
goes back to itself up to a symmetry transformation $U_\beta$:
$\psi(y+ 2 \pi \rho)=U_\beta \psi(y)$. In particular, if the twist matrix is 
different for bosons and fermions, 
such boundary conditions lead to supersymmetry breakdown. If the extra dimension is an orbifold, say $S^1/\Gamma$ where $\Gamma$ is a discrete group, for instance $Z_2$ or 
$Z_2\times Z_{2}'$, then $\Gamma$ may be embedded nontrivially into the group 
of global and local symmetries of the model, and its combined geometrical 
and internal action has to be taken into account. 
Finally, the matter and gauge fields located on the branes have in general 
nontrivial interaction with bulk fields, and bulk fields must be allowed 
to have selfinteractions. In the context of a locally supersymmetric theory 
bulk selfinteractions imply that one has a gauged supergravity in the bulk,
with a number of hypermultiplets and vector and tensor multiplets. 
The coupling of gr avity and and bulk matter to branes implies that branes act as nontrivial sources for bulk configurations, and modify the behaviour of 
fields at the fixed points. We shall try to determine how the Scherk-Schwarz
mechanism works in the presence of all these ingredients and how does it 
affect stability of the extra dimensions. 
\section{Flipped and detuned supergravity in five dimensions}
The simple N=2 d=5 supergravity multiplet contains metric tensor (represented by 
the vielbein $e^m_\alpha$), two gravitini $\Psi^A_\alpha$ and one vector field $A_\alpha$ -- the graviphoton. We shall consider gauging of a ($U(1)$) subgroup of the global $SU(2)_R$ symmetry of the 5d Lagrangian. In general, coupling of bulk fields to branes turns out to be related to the gauging, and the bulk-brane couplings will preserve only a subgroup of the $SU(2)_R$. 
Purely gravitational 5d action describing such a setup reads $
%        \begin{equation}
        S=\int_{M_5}\ e_5 \ {\cal L}_{grav}\ , 
%\label{dzialabiesugra}
%        \end{equation} 
$        where
        \begin{eqnarray} 
        &{\cal L}_{grav}=&\frac{1}{2}R-\frac{3}{4}{\cal F}_{\alpha\beta}{\cal F}^{\alpha\beta}-\frac{1}{2\sqrt{2}}A_\alpha{\cal F}_{\beta\gamma}{\cal F}_{\delta\epsilon}\epsilon^{\alpha\beta\gamma\delta\epsilon}+\nonumber\\&&-\frac{1}{2}\bar{\Psi}^A_\alpha\gamma^{\alpha\beta\gamma}D_\beta\Psi_{\gamma A}+\frac{3{\rm i}}{8\sqrt{2}}\left(\bar{\Psi}^A_\gamma\gamma^{\alpha\beta\gamma\delta}\Psi_{\delta A}+2\bar{\Psi}^{\alpha A}\Psi_{A}^{\beta}\right){\cal F}_{\alpha\beta}+\nonumber\\&&-\frac{{\rm i}}{\sqrt{2}} {\cal P}_{AB}\bar{\Psi}^A_\alpha\gamma^{\alpha\beta}\Psi_{\beta}^B-\frac{8}{3} {\rm Tr}({\cal P}^{2})\ .
        \end{eqnarray} 
        Covariant derivative contains both gravitational and gauge connections:
        \begin{equation} 
        D_\alpha\Psi_\beta^A=\nabla_\alpha\Psi_\beta^A+ A_\alpha{\cal P}^A_B\Psi_\beta^B\ ,
        \end{equation} 
where $\nabla_\alpha$ denotes covariant derivative with respect to gravitational transformations and ${\cal P}={\cal P}_i \, {\rm i}\, \sigma^i$ is the 
gauge prepotential. The pair of gravitini satisfies symplectic Majorana condition
 $\bar{\Psi}^A\equiv\Psi_A^\dagger\gamma_0=(\epsilon^{AB}\Psi_B)^TC$ where $C$ is the charge conjugation matrix and $\epsilon^{AB}$ is antisymmetric $SU(2)_R$ metric (we use $\epsilon_{12}=\epsilon^{12}=1$ convention). 
Supersymmetry transformations are also modified by the gauging
        \begin{eqnarray} 
        &&\delta e^m_\alpha=\frac{1}{2}
\bar{\epsilon}^A\gamma^m\Psi_{\alpha A},\;\;\delta A_\alpha=\frac{{\rm i}}{2\sqrt{2}}\Psi_{\alpha}^A\epsilon_A,\\&&\delta\Psi_{\alpha}^A=D_\alpha\epsilon^A- \frac{{\rm i}}{6\sqrt{2}}\left(\gamma_\alpha^{\beta\gamma}-4\delta_\alpha^\beta\gamma^\gamma\right){\cal F}_{\beta\gamma}\epsilon^A+\frac{\sqrt{2}{\rm i}}{3}{\cal P}^{AB}\gamma_\alpha\epsilon_B\ .
        \end{eqnarray} 

%\end{document}

If one puts ${\cal P}=0$ and stays on the circle, then as the twist matrix 
one may take any $SU(2)$ matrix acting on the symplectic indices $a=1,2$. 
On a circle the $U(1)$ prepotential takes the form $ P = g_S s_a \, i \, \sigma^a$ and the twist matrix is $U_\beta = e^{i\, \beta \, s_a \sigma^a}$. However, in this case the unbroken symmetry is a local one, and the Scherk-Schwarz condition is equivalent to putting in a nontrivial Wilson line, \cite{vonGersdorff:2002tj}, we shall come back to this issue later in the paper. 

When one moves over to an orbifold $S^1 / \Gamma$, one needs to define in addition to the gauging the action of the space group $\Gamma$ on the fields. 
Let us take $\Gamma = Z_2$ first. Then we have two fixed points at $y=0,\pi$, 
and we can define the action of $Z_2$ in terms of two independent boundary 
conditions ($\Psi$ stands here for a doublet of symplectic-Majorana spinors or for a doublet of scalars, like two complex scalars from the hypermultiplet)   
\begin{equation} \label{warunkiSS}
        \Psi(-y)={\hat{Q}}_0\Psi(y)\ ,\quad\Psi(\pi r_c-y)={\hat{Q}}_\pi\Psi(\pi r_c+y)\ ,
        \end{equation}
        where $\hat{Q}_0$, $\hat{Q}_\pi$ are some arbitrary matrices, independent of the space-time coordinates, such that  ${\hat{Q}}_0^2={\hat{Q}}_\pi^2=1$ . Conditions (\ref{warunkiSS}) imply:
        \begin{equation}
        \Psi(y+2\pi r_c)={\hat{Q}}_\pi {\hat{Q}}_0\Psi(y)\ .
        \end{equation}
Hence, if the boundary conditions at $y=0$ and $y=\pi r_c$ are different, 
one obtains twisted boundary conditions with $U_\beta ={\hat{Q}}_\pi {\hat{Q}}_0$.
It is easy to see that $U_\beta \hat{Q}_{0,\pi}  U_\beta= \hat{Q}_{0,\pi}$,
which is the consistency condition considered in \cite{Bagger:2001ep},\cite{Biggio:2002rb},\cite{Meissner:2002dg}.
This is immediately  generalized to $S^1/(Z_2 \times Z_2')$ with two fixed points 
for each of the $Z_2$s, $y=0,\frac{1}{2}\pi r_c, \pi r_c, \frac{3}{2} \pi r_c$, and independent 
$\hat{Q}_y$ at each of the fixed points.  

If one writes $\hat{Q}_\pi \hat{Q}_0=\exp(i\beta_a\sigma^a)$, the condition (\ref{warunkiSS}) is solved by
        \begin{equation}
        \Psi=e^{i\beta_a\sigma^a f(y)}\hat{\Psi}\ ,
        \end{equation}
        where $\hat{\Psi}$ is periodic on the circle and $f(y)$ obeys the conditions
        \begin{equation}
        f(y+2\pi r_c)=f(y)+1\ ,\quad f(-y)=- f(y)\ .
        \end{equation}
        
When one expresses the initial fields $\Psi$ through $\hat{\Psi}$, the kinetic term in the Lagrangian generates  mass terms for {\em periodic} fields $\hat{\Psi}$:
        \begin{equation}
        \bar{\Psi}\gamma^M\partial_M\Psi\supset if'\bar{\hat{\Psi}}\gamma^5\beta_a\sigma^a\hat{\Psi} \ .
        \end{equation}
        \vskip0.3cm
        
%\end{document}
\subsection{Scherk-Schwarz mechanism in the $SU(2)$ R-symmetry of 5d gauged 
supergravity}

Let us now move on to the specific case of a 5d supergravity with a 
gauged $U(1)$ subgroup of the $SU(2)$ R symmetry. 
%We will consider case with fifth dimension as an orbifold, which is interesting f%rom Randall-Sundrum scenario point of view. 
The $Z_2$ action on the gravitino is defined as follows:
\begin{eqnarray} \label{gbcond}
        &\Psi^A_\mu(-y)=\gamma_5(Q_0)^A_B\Psi^B_\mu(y)\ ,\quad\Psi^A_5(-y)=-\gamma_5(Q_0)^A_B\Psi^B_5(y)\ ,&\nonumber\\&\Psi^A_\mu(\pi r_c-y)=\gamma_5(Q_\pi)^A_B\Psi^B_\mu(\pi r_c+y)\ ,\quad\Psi^A_5(\pi r_c-y)=-\gamma_5(Q_\pi)^A_B\Psi^B_5(\pi r_c+y),&
\end{eqnarray}
and the parameters $\epsilon^A$ of the supersymmetry transformations obey the same boundary conditions as the 4d components of gravitini. 
%\end{document}
Symplectic Majorana condition 
($(Q_{0,\pi})^C=\sigma_2 (Q_{0,\pi})^{*}\sigma_2=-Q_{0,\pi}$) and normalization 
$(Q_{0,\pi})^2=1$ imply $Q_{0,\pi}=(q_{0,\pi})_a \sigma^a$, where $(q_{0,\pi})_a$ are real parameters \cite{Bergshoeff:2000zn}.
%\end{document}
We would like to gauge a $U(1)$ subgroup of the global $SU(2)$. 
In the general case \cite{Brax:2001xf} we can choose the prepotential 
of the form
\begin{equation} 
\label{prep}
        P = g_R \epsilon(y) R + g_{S} S,
\end{equation}
%\end{document}
where $R=r_a i \sigma^a$ and  $S=s_a i\sigma^a$. On an orbifold 
$S^1/(Z_2 \times Z_2')$ the expression $\epsilon(y)R$ gets replaced by 
$\bar{R}(y)$ which is a pice-wise constant matrix with discontinuities (jumps) at the positions of the four branes. The basic relation between the boundary conditions and the prepotential comes from the requirement, that under supersymmetry variations the 
transformed gravitino $\Psi^{A}_\alpha + \delta \Psi^{A}_\alpha$ 
should obey the same boundary conditions as $\Psi^{A}_\alpha$. Taking into account that the gauge field present in the supersymmetry transformation of the gravitini is that graviphoton, whose 4d part we choose to take $Z_2$-odd with respect to each brane (we need only $N=1$ supersymmetry on the branes), and the fifth component is always even, we obtain the relations  valid for any segment containing a pair of naighbouring fixed points
\begin{equation} 
\label{barel}
 [ Q_{0, \pi}, R  ] = 0, \;\;  \{  Q_{0, \pi}, S  \} = 0 \, .
\end{equation}
For nonzero $R$ this implies $Q_y$ proportional to $R$, i.e. 
$Q_{y}= \alpha \, ( i \, \sqrt{R^2})^{-1} \, R$ with $\alpha= \pm 1$. The simplest case  
of interest corresponds to $Q_0 = - Q_\pi $. As shown in \cite{Brax:2001xf}, in this case the closure of supersymmetry transformations reqires putting on the branes  equal tensions whose maginitude is determined by $R$
(we quote only the bosonic gravity part of the action):
%end{document}
\begin{equation}
M^{-3} S=\int d^5 x \sqrt{-g_5} (\frac{1}{2}R + 6 k^2)-  6 \int d^5 x\sqrt{-g_4}k T (\delta(x^5) + \delta(x^5-\pi r_c))
\end{equation}
%end{document}
where 
\begin{equation}
k = \sqrt{ \frac{8}{9}(g_R^2 R^2+ g_S^2 S^2)} \; {\rm and} \; 
%\end{equation}
% and
%\begin{equation}
T= \frac{g_1\sqrt{\vec{R}^2}}{\sqrt{(g_1^2 R^2+ g_2^2S^2)}}.
\end{equation}
This is easily generalized. 
If on a $S^1/(\Pi\, Z_2)$ one takes boundary conditions given by pairs of $Q$ and $-Q$ one after another, then this implies that all branes 
on $S^1/Z_2$, $S^1/(Z_2 \times Z_2')$, $S^1/(\Pi\, Z_2)$ have the same brane tension. 
Assuming also $S \neq 0$ such a system gives a static vacuum with $AdS_4$ 
foliation and fixed radius of the orbifold\footnote{For two equal and negative brane tensions the maximally symmetric solution doesn't exist.}. In the case of $Z_2$ the overall twist matrix is given by $U_\beta = - {1} $ and in the case of $Z_2 \times Z_2'$
there is no overall twist: $U_\beta= + {1}$. This may be
generalized again. From the analysis of \cite{Brax:2001xf} it follows that if 
in the boundary conditions $Q$ is followed by $+Q$ (and not $-Q$) on the next 
brane, then the brane tension on the second brane must be equal in magnitude but of opposite sign to that on the first brane. Together with the previous findings this leads to quasi-quiver diagrams where branes with brane tensions 
$\pm \lambda$ 
and boundary conditions $(\pm Q),(\pm Q)...$ follow each other respecting 
$\Pi Z_2$ symmetry. At first glance possible strings of boundary conditions could be for instance $(Q),(Q),(-Q),(-Q)$ or $(Q),(-Q),(-Q),(Q),(-Q),(-Q)$. 
However, the stretches between $(Q),(Q)$ branes collapse to a point. 
This is easily seen from the fact, that to have a finite-length distance 
between $(-Q),(+Q)$ branes the brane tensions must be smaller than 
the critical value $\lambda < \lambda_{cr}$, equivalently $T < 1$,
where the critical tension is that of the Randall--Sundrum brane. 
This implies, that the branes at the ends of a $(Q),(Q)$ segment 
have brane tensions of opposite sign and of the same magnitude smaller than 
the critical value. Soving boundary conditions on such a segment, with the 
maximally symmetric foliation ansatz, leads to the result $|y_{b1}-y_{b2}|=0$,
i.e. the branes coincide. Hence the possible static quasi-quivers 
correspond to the chains $(Q),(-Q),(Q),...,(-Q),(Q),(-Q)$. These are 
locally supersymmetric backgrounds corresponding to the models of the type 
discussed in \cite{Barbieri:2000vh}.  

Another extension of the picture is possible.  
Instead of cancellation of the supersymmetry variation of the brane tension 
by means of a `jumping' prepotential in the bulk, one can consider, 
\cite{Altendorfer:2000rr},\cite{Bagger:2002rw},
adding 4d gravitino mass terms on branes, and modifying the variations of 
the fifth component of the gravitino. Also in this case one can impose 
different boundary conditions on different walls. 
The general case of such an extension shall be discussed elsewhere, 
here we quote a simple example with $R=0$ for illustration. In this case one has a brane action of the form
\begin{equation} \label{second}
S_b = \int d^5x \, e_4 \delta(y-y_b) ( - \lambda_0 + M_{AB} \bar{\Psi}^{A}_{\mu}
\gamma^{\mu \nu} \Psi^{B}_\nu) 
\end{equation}
with gravitini components $\Psi^{B}_\nu$ satisfying local boundary conditions
given by a matrix $Q_{y_b}$. The necessary modification of the transformation law 
is $\delta \Psi^{A}_{\hat{5}} =  2 \delta(y-y_b) \epsilon^{AC} M_{(CB)} \gamma_5 
\epsilon^{B}$ where the hat denotes a flat space index and $(..)$ is the symmetrization with the weight $1/2$. Cancellation of susy variations requires that 
the brane tension and the matrix $M_{AB}$ satisfy the equation
\begin{equation}
\lambda_0 \epsilon_{AB} + 4 \sqrt{2} \, i \, ( M_{(AC)} P^{C}_{B} + P_{(AC)} \epsilon^{CD} M_{(DB)} )=0 \, .
\end{equation} 
In the above suitable projections are made with the help of the projective operators $\Pi_{\pm} = 1/2 (\delta^{A}_{B} \pm \gamma_5 {Q_{y_b}}^{A}_{B})$, remembering that $\Psi^{A}_5 $ and $\Psi^{B}_{\mu}$  have opposite parities, see 
(\ref{gbcond}). In particular, the simple choice $\lambda_0 = M_{AB} = 0$
is sufficient to illustrate the way the twisting works. In this case   
the relevant condition on the operators $Q_{y_b}$ is $\{ Q_{y_b}, P \} =0$. 
 
Let us discuss the genaration of the Scherk-Schwarz (nonsupersymmetric) 
mass terms in the two limiting cases. First, let us take $\Gamma=Z_2$ and 
{ $Q_{0}$ and $Q_{\pi}$ orthogonal to each other}. 
In this case one is free to take $Q_{0}=\sigma_3$ and $Q_{\pi}=\sigma_1$. 
Then $Q_{\pi}Q_{0}=-i\sigma_2$. Hence, the only possibility for the prepotential is $gP= ig\sigma_2$, because other directions do not commute or anticommute with both $Q_{0}$ and $Q_{\pi}$.
In such case the only unbroken $U(1)$ subgroup is generated by the  $\sigma_2$ direction of the global $SU(2)$ symmetry. Notice that we can write $Q_{\pi}Q_{0}=-i\sigma_2=\exp(i\beta_2\sigma^2)$, where $\beta_2=\frac{3}{2}\pi+2k\pi$ and $k\in Z$. Then the solution which satisfies the  boundary conditions, expressed in terms of periodic fields, is:
        \begin{equation} \label{firstex1}
        \Psi=e^{i\beta_2\sigma^2 f(y)}\hat{\Psi}\ .
        \end{equation} 
Hence in the action expressed in terms of periodic fields, one obtains supersymmetry violating mass terms:
        \begin{equation} \label{firstex2}
-e_5 \frac{{\rm i}}{2} {\delta}_{AB}\bar{\Psi}^A_\mu\gamma^{\mu\nu} \gamma^5\Psi_{\nu}^B \, \beta_2 \, f'(y).
        \end{equation}
Let us take as an example the function $f=y/(2 \pi r_c)$. This function 
leads to the mass term $-\frac{{\rm i}}{2} {\delta}_{AB}\bar{\Psi}^A_\mu\gamma^{\mu\nu} \gamma^5\Psi_{\nu}^B \frac{\beta_2}{2 \pi r_c}$.
Already here one can conclude that the bulk Lagrangian after redefinition cannot be put into the form compatible with linearly realized supersymmetry. To see this, one should note that the only mass terms compatible with supersymmetry are given by  a prepotential. 
Supersymmetry requires, that the same prepotential determines the bulk scalar potential. In our case, we have redefined the gravitini only, hence the 
bulk potential term stays unchanged, independent of $\beta_2$. 
At the same time any prepotential that should describe the Scherk-Schwarz 
mass terms shall depend on $\beta_2$, hence the supersymmetric 
relation between 
mass terms and the scalar potential is necessarily violated. This is what we mean when we call the Scherk-Schwarz mass terms explicitly non-supersymmetric. 
On the other hand it is obvious, that the Scherk-Schwarz picture is 
equivalent indeed to a spontaneosly broken flipped supergravity.   

Let us now consider again the case where { $Q_{0}$ and $Q_{\pi}$ are parallel.} Let us take for simplicity $Q_{0}=\sigma_3$. Then $Q_{\pi}=\alpha\sigma_3$, where $\alpha=\pm 1$, and the twisted boundary conditions take the form:
        \begin{equation}
        \Psi(y+2\pi r_c)=\alpha\Psi(y)\ .
        \end{equation} 
        For $\alpha=1$ we have usual case with periodic field. For $\alpha=-1$ we obtain `flipped' supersymmetry of \cite{Brax:2001xf}. Let us take a nonzero $S$-part of (\ref{prep}). 
%$S=0$ is not interesting for us, because there is not stable vacuum in such case %\cite{brax}. 
Assume the prepotential of the form
        \begin{equation} 
        P=\frac{g}{\sqrt{2}}\left(\epsilon(y)\sigma_3+\sigma_1\right)\ .
        \end{equation} 
        For $\alpha=-1$ we can write:
        \begin{equation} 
        Q_{\pi} Q_0=-1=e^{i\beta(\epsilon(y)\sigma_3+\sigma_1)}\ ,
        \end{equation} 
        where $\beta=\pi+2k\pi$ and $k\in Z$. Similarly to the case of the previous paragraph one obtains the following solution
        \begin{equation} 
        \Psi=e^{i\beta(\epsilon(y)\sigma_3+\sigma_1) f(y)}\hat{\Psi}\ ,
        \end{equation} 
        and supersymmetry violating mass terms
        \begin{equation} 
-e_5 \frac{{1}}{2} {\epsilon}_{AB}\bar{\Psi}^A_\mu\gamma^{\mu\nu} \gamma^5
\left ( \int_{0}^{1} ds e^{i s \beta ( \epsilon(y) \sigma_3 + \sigma_1)}
( {\rm i } \beta (\epsilon f)' \sigma_3 + {\rm i } \beta f' \sigma_1 ) 
e^{ i (1-s)\beta ( \epsilon(y) \sigma_3 + \sigma_1)} \right )^{B}_{C} 
\Psi_{\nu}^C \,.
\end{equation} 
Let us take again $f=y/(2 \pi r_c)$. This gives 
\begin{eqnarray} 
&-e_5 \frac{{1}}{2} \bar{\Psi}^A_\mu\gamma^{\mu\nu} \gamma^5
\frac{\beta}{2 \pi r_c} ( \sigma^{3}_{AB} - \epsilon(y) \sigma^{1}_{AB})
\Psi_{\nu}^B  \, - e_4 \delta(y-\pi r_c) \bar{\Psi}^A_\mu
\gamma^{\mu \nu} \gamma^{\hat{5}}  \sin(\beta/2)  
 \sigma^{1}_{AB} \Psi_{\nu}^B & \nonumber \\
&= -e_5 \frac{{1}}{2} \bar{\Psi}^A_\mu\gamma^{\mu \nu} \gamma^5
\frac{1}{2 r_c} ( \sigma^{3}_{AB} - \epsilon(y) \sigma^{1}_{AB})
\Psi_{\nu}^{B} - e_4 \delta(y-\pi r_c) \bar{\Psi}^{A}_\mu \gamma^{\mu\nu} 
\gamma^{\hat{5}}  \sigma^{1}_{AB} \Psi_{\nu}^{B}.& 
\end{eqnarray} 
This example is more involved, but the same comments as in the previous case 
apply. 
The generalization to a quasi-quiver setup is obvious. 
One may notice, that only the bulk terms are proportional to the naive 
KK scale $1/r_c$. The scale of boundary terms is set by the 5d Planck scale. 

Let us note already here, that even though the symmetry that we are using 
to implement the Scherk-Schwarz mechanism may be a local one, the Scherk-Schwarz masses cannot be removed, as one may naively think, by means of a gauge transformation. Such a transformation would have to be a `large' one, leading from a periodic to an antiperiodic configuration. However, the definition of the model involves not only couplings in the Lagrangian but also the choice of specific boundary conditions. Hence such large gauge transformations connect two different (although physically equivalent) Hilbert spaces, and do not belong to the group of internal symmetries of our models.

We would like to note, that in a curved gravitational background 
different mass spectra for the `would be' superpartners, like graviton and gravitino, is not an unambigous sign of broken supersymmetry. In particular, in flipped supersymmetry models the background is of the $AdS_4$-foliation form, and one knows \cite{Heidenreich:rz},\cite{Ferrara:1998ej} that the $AdS_4$ supermultiplets contain in general particles with different mass terms. For instance, massive higher spin representations ($E_0 
> s+1, \, s \geq 1/2$) are of the form 
\begin{equation}
D(E_0,s) \oplus D(E_0 +1/2, s+ 1/2) \oplus D(E_0 + 1/2, s-1/2) \oplus D(E_0 +1, s),
\end{equation} 
with the mass-squared operator $m^2= E_0 (E_0 -3) - (s+1)(s-2)$.
However, if one is able to obtain an approximate formula 
for a mass spectrum as a function of the quantization parameter, then 
one can compare its shape to the towers of supersymmetric masses. 
Since in any case the mass terms are certainly of phenomenological interest, we shall compute them and come back to the issue of seeing supersymmetry breaking
through the spectrum later in the paper. 
The unambigous  sign of supersymmetry breakdown are  nonzero vacuum values of the variations of fermions, or the absence of global Killing spinors, 
see \cite{Brax:2001xf}. 
    
\subsection{Another picture of the Scherk-Schwarz mechanism in the presence of 
gauge symmetries}
Let us consider again the 5d supergravity with $U(1)$ gauge symmetry and charged gravitno fields (i.e. $U(1)$ subgroup of $SU(2)_R$). 
The covariant derivative takes the form 
$D_M\Psi^{A}_N=\nabla_M\Psi^{A}_N+gA_M P^{A}_{B}\Psi^{B}_N$ .
In the case of unbroken supersymmetry all components of
gauge fields should have vacuum expectation values equal to zero. 
To see this, one may have a look at  the gravitino supersymmetry 
transformation, $\delta_\epsilon\Psi^A_M\supset\partial_M\epsilon^A+gA_M P^A_B\epsilon^B$. 
In the previous section we have assumed the expectation value of the gauge field to vanish, $<A_M>=0$, but the gravitini fields were choosen to satify 
twisted boundary conditions.  
On the other hand, one may try to eliminate twisted boundary conditions by a 
non-periodic (large) gauge transformation. Let us take for the sake of 
definiteness the first of the examples of the previous section, equations 
(\ref{firstex1}),(\ref{firstex2}), $P=i g \sigma^2$. 
The necessary gauge transformation is given by 
        \begin{equation}
        \Psi_M\rightarrow \Psi'_M = e^{-{\rm i} \beta_2 \sigma^2 f(y)}\Psi_M\ ,
        \end{equation} 
and obviously the primed gravitino field is periodic. 
However in such a case one has to 
transform the gauge field as well
        \begin{equation} 
        \tilde{A}_5\rightarrow \tilde{A}'_5= \tilde{A}_5 + {\rm i} \beta_2 \sigma^2 f'(y),
        \end{equation} 
where the tilde denotes matrix gauge field. This implies  
in turn  a non-zero vacuum expectation value of the transformed $U(1)$ 
gauge field $<A'_5>=1/g f'(y)\beta_2$. At present the bulk 
non-supersymmetric gravitino masses arises in the theory as a result of 
a non-zero expectation value of the gauge part of the covariant derivative in the kinetic term.
This is the Wilson line breaking, see \cite{vonGersdorff:2002tj}. 
Let us note that if one takes our standard choice $f(y)= y/(2 \pi r_c)$, then  the integrated Wilson line does not vanish, 
$\oint dx^5 <A_5> = {\rm i} \beta_2 \sigma^2 \neq 0$, 
which is a sign that supersymmetry is broken globally. 

\section{The role of boundary couplings}

The universal hipermultiplet $\{\zeta^a,q^i\}$ consists of a doublet of fermions 
and of four real scalars $q_i\in\{V,\sigma,x,y\}$ (we also use $\xi=x+{\rm i}y$). 
Scalars parametrize a quaternionic manifold, whose global symmetry group is 
$Sp(2)\times SU(2)_R$. The kinetic term of the bosonic part reads:
        \begin{equation} 
        \scr{L}_{kinetic}=-h_{ij}D_M q^i D^M q^j\ ,\quad h_{ij}=V^{Aa}_iV^{Bb}_j\epsilon_{AB}\Omega_{ab}\ ,
        \end{equation}   
        where $h_{ij}$ is a quaternionic metric. 
%$V^{Aa}_i$ associated vielbein and $\Omega_{ab}$, $\epsilon_{AB}$ are antisymmetric $Sp(2)$, $SU(2)_R$ metrics respectively (we use $\Omega_{21}=\Omega^{21}=\epsilon_{12}=\epsilon^{12}=1$ conventions). 
We can write the quaternionic metric in the explicit form:
        \begin{equation} 
        h_{ij} dq^idq^j=\frac{1}{4V^{2}}dV^{2}+\frac{1}{4V^{2}}\left[d\sigma+{\rm i}(\bar{\xi}d\xi-\xi d\bar{\xi})\right]^{2}+\frac{1}{V}d\xi d\bar{\xi}\ .
        \end{equation} 
%        and vielbein
%        \begin{equation} 
%        V^{Aa}=V^{Aa}_idq^i=\frac{1}{\sqrt{2}}\left(\begin{array}{cc} u&\bar{v}\\v&-\bar{u}\end{array}\right)^{Aa}\ ,
%        \end{equation} 
%        where
%        \begin{equation} 
%        u=\frac{d\xi}{\sqrt{V}}\ ,\qquad v=\frac{1}{2V}\left(dV+{\rm i}d\sigma+\xi d\bar{\xi}-\bar{\xi}d\xi\right)\ .
%        \end{equation} 
As explained in \cite{Falkowski:2000er},\cite{Falkowski:2000yq}, when some of the global  symmetries are gauged, 
supersymmetry requires additional boundary terms involving bulk fields to be 
present in the model. This is the way the five-dimensional 
version 
of the Horava-Witten model, and brane-bulk supergravities of \cite{Falkowski:2000er},\cite{Bergshoeff:2000zn} work.  
{}For simplicity let us consider again the $Z_2$ orbifold (generalizations 
follow the same lines as in the previous chapters). 
In the case of flipped Horava-Witten and Randall-Sundrum models (with the hypermultiplet) the additional boundary terms in the action are 
\begin{equation}
S_b= - \int d^5 x e_4 6 k (\theta - \frac{\alpha}{V} - \theta \frac{|\xi|^2}{V})
(\delta(y) + \delta(y-\pi r_c)) \, ,
\end{equation}  
where $\theta=0$ gives the flipped Horava-Witten model, and $\alpha=0$ corresponds to
the flipped Randall-Sundrum case. These terms, like the gauging in the bulk, 
preserve only a $U(1)$ subgroup of the product of the $SU(2)_R$ and the symmetry group of the quaternionic manifold. Moreover, these boundary terms break explicitly 
the $N=2$ bulk supersymmetry down to the local $N=1$. 
It has been shown in \cite{Dudas:1997jn} that if one would have at ones disposal the 
exact $SU(2)_R$ symmetry and embed the twist of the boundary conditions into this group to break $N=2$ supersymmetry by the Scherk-Schwarz mechanism in the Horava-Witten model, and in addition project-out the $Z_2$-odd mode, then in four dimensions this breaking would be seen as a spontaneous breaking of the efective $N=1$ supergravity. However, the complete model contains also 
brane terms that break $N=2$ explicitly. Second, as we have seen having a net twist
of the boundary conditions requires opposite boundary conditions on each wall, 
and the same sign of boundary terms on both. This makes a physical difference 
with respect to the original Horava-Witten model, where the boundary terms had to have opposite signs. These observations imply in particular, that 
the five-dimensional picture of the Scherk-Schwarz supersymmetry breaking in 
Horava-Witten and Randall-Sundrum type models is physically different from the 
breaking by gaugino or superpotential (or flux) condensation in these models,
see for instance 
\cite{Lalak:1997zu},\cite{Falkowski:2000er},\cite{Ellis:1998dh}. 

To illustrate how the Scherk-Schwarz breaking would work in a model with a hypermultiplet, let us put $\theta=0$. One can check that in this case the full action 
(including boundary terms) is invariant under the following transformation:
        \begin{equation} 
        \zeta^a\rightarrow e^{-{\rm i}\beta\sigma_3}\zeta^a\ ,\quad\Psi^A\rightarrow e^{-{\rm i}\beta\sigma_3}\Psi^A\ ,\quad\xi\rightarrow e^{{\rm i}2\beta}\xi\ ,
        \end{equation}
        where $\beta$ is the transformation parameter and $\zeta^a$ is a hyperino.
Now, one can impose 
associated boundary conditions:
        \begin{equation}\label{warunkihiperss}
        \zeta^a(y+2\pi r_c)=e^{-{\rm i}\beta\sigma_3}\zeta^a(y)\ ,\quad\Psi^A(y+2\pi r_c)= e^{-{\rm i}\beta\sigma_3}\Psi^A(y)\ ,\quad\xi(y+2\pi r_c)=e^{{\rm i}2\beta}\xi(y)\ .
        \end{equation}
        It is obvious that fields
        \begin{equation}
        \zeta^a=e^{-{\rm i}\beta\sigma_3f(y)}\hat{\zeta}^a\ ,\quad\Psi^A= e^{-{\rm i}\beta\sigma_3f(y)}\hat{\Psi}^A\ ,\quad\xi=e^{{\rm i}2\beta f(y)}\hat{\xi}
        \end{equation}
        expressed in terms of new periodic (hatted) fields satisfy the conditions 
(\ref{warunkihiperss}). As the consequence the Scherk-Schwarz 
non-supersymmetric mass terms are generated for hyperini
\begin{equation}
\frac{1}{2} e_4 ({\rm i} \beta) \bar{\zeta}^a \gamma^{\hat{5}} \sigma^{1}_{ab} \zeta^b f'(y),
\end{equation} 
gravitini, and for the complex scalars $\xi$. In fact, one generates also new 
quartic terms in the scalar potential, in addition to the supersymmetric 
scalar potential $V_{susy}=g^2 8/3 tr (P^2) + g^2/2 h_{ij} k^i k^j$, where 
here $k^{\xi}= 2 {\rm i} \xi$. 
Assuming vanishing expectation values of brane sources for the scalar 
$\sigma$ the terms in the non-supersymmetric part of the potential that do 
not contain $\partial_5 \xi$ are
\begin{equation} 
V(\hat{\xi}) = \frac{\beta^2 (f')^2}{V^2} |\hat{\xi}|^2 ( 4 V + 
3 |\hat{\xi}|^2),
\end{equation}  
hence the mass-squared parameter for the normalized scalar is 
$4 \beta^2 (f')^2$. 
One can see that performing the Scherk-Schwarz redefinition 
of scalar fields, which my be identified for instance with the higgs-like 
field in the observable sector, one can create a complicated scalar 
potential. However, it will always contain the same physics as the locally 
supersymmetric lagrangian in the spontaneously broken phase, which is usually 
much simpler to analyse. The same comment concerns the fermionic sector. 
The Scherk-Schwarz masses for matter fermions 
superficially look like terms breaking 
supersymmetry in a hard way (like quartic terms in the potential), but 
the equivalence to the spontaneously broken phase guarantees cancellation of 
dangerous divergencies.  
The fact that the Scherk-Schwarz masses for chiral fermions do not belong to a 
linearly realized 5d supersymmetry may be seen from the observation, that 
supersymmetric masses are defined by the geometry of the quaternionic manifold 
and by the Killing vectors $k^i$, none of which had changed under the 
Scherk-Schwarz redefinition. 
To summarize, the redefinitions have broken linear supersymmetry both in 
hipermultiplet and in gravity sectors.

%%%%%%%%%%%%%%%%%%%%%%%%%%%%%%%%%%%%%%%%%%%%%%%%%%%%%%%%%%%%%%%%%%%%%%%%%%%

Another issue concerning boundary couplings of bulk fields is a statement, 
that they may lead to dangerous singular terms in the equations of motion of the bulk fields, and therefore one should use various redefinitions of fields to redefine them away. In fact, in consistent theories such as bulk--branes supergravities 
these singularities are harmless, see \cite{Ellis:1998dh},\cite{Lalak:2001fd}. 
To see how the cancellation works, let us take the example of a $Z_2$-odd bulk 
field coupled derivatively to brane operators, which is of particular importance in 
Horava-Witten and Randall--Sundrum type models. 
In this case the Lagrangian includes a coupling to sources on the
hidden wall
and to operators consisting of observable fields on the visible wall 
\beqa
S(\Phi) & = & \int d^5 x \left ( \frac{1}{2} \pa \Phi \pa \Phi +
\pa_5 \Phi {\cal S} \delta(x^5-\pi \rho) + \pa_5 \Phi {\cal O} \delta(x^5) \right )
\eeqa
The bulk equation of motion is 
\begin{equation}
 \Box_4 \Phi + \pa_5^{2} \Phi = \cs (x) \pa_5 \delta(x^5 -\pi \rho) + 
\co (x) \pa_5 \delta(x^5)
\end{equation}
The equation of motion for the non-zero mode $ \psi (x;x^5)$ coincides then
with the full equation of motion, 
the proper boundary conditions on the half-circle being 
\beqa 
{}&{} & \lim_{x^5 \rightarrow \pi \rho} \psi= 
\frac{1}{2} \cs  
\nonumber \\ 
{}&{} & \lim_{x^5 \rightarrow 0} \psi = 
- \frac{1}{2} \co 
\eeqa
One easily finds the solution in the form 
\begin{equation}
\psi = \frac{\cs + \co}{2\pi \rho} x^5 - \frac{\co}{2} + \psi_1 + \psi_2 +
\ldots
\label{e66}
\end{equation}
where the higher terms in the series are vanishing on the branes and 
can be computed from the recursive relation 
$ \Box_4 \psi_{n-1} + \pa_5^{2} \psi_n =0$. 
One can check, that the sigular terms cancel out from the equations of motion of 
all fields $\psi_n$, $n=1,2,...$. For instance, the equation of motion for $\psi_1$
is 
\begin{equation}
 \Box_4 \psi_1 + \pa_5^{2} \psi_1 = \frac{ -\Box_4 ( \co + \cs) }{2 \pi r_c} y
 + \frac{\Box_4 \co}{2}.
\end{equation}
Hence the field $\psi_1$ has no discontinuities, and no singularities 
in the equation of motion. It is interesting to notice, that derivatives of boundary operators act effectively as bulk sources for $\psi_1$. 

\section{Wave functions and mass quantization in flipped supergravity}% (bi-supergravity)}

In this chapter we would like to have a closer look at the localization of wave functions and mass quantization in simple models with twisted supersymmetry, and compare this to the well known Randall-Sundrum case with supersymmetry and 
to the same model with supersymmetry explicitly broken through the `wrong'
sign of the brane tension on one of the walls. The specific twisted model we 
shall discuss here is the locally supersymmetric generalization of the 
$(++)$ bigravity model of \cite{Kogan:2000vb}.  
 
\subsection{Naive RS model with flipped boundary conditions} \label{pomarolr}
        
Let us focus on the supergravity action  with the prepotential of the form $P=g\epsilon(y){\rm i}\sigma_3R$. We define $Z_2$ action on the gravitino sector as:
        \begin{eqnarray} \label{plaskiz2}
        &\Psi^A_\mu(-y)=\gamma_5(\sigma_3)^A_B\Psi^B_\mu(y)\ ,\quad\Psi^A_5(-y)=-\gamma_5(\sigma_3)^A_B\Psi^B_5(y)\ ,&\nonumber\\&\Psi^A_\mu(\pi r_c-y)=-\gamma_5(\sigma_3)^A_B\Psi^B_\mu(\pi r_c+y)\ ,\quad\Psi^A_5(\pi r_c-y)=\gamma_5(\sigma_3)^A_B\Psi^B_5(\pi r_c+y)&\ ,
        \end{eqnarray}
        which implies flipped ($\Psi_\alpha^A(y+2\pi r_c)=-\Psi_\alpha^A(y)$) boundary conditions. Cosmological constant that arises from the prepotential is given by $\Lambda_5=-\frac{16}{3}g^2 R^2$. To obtain the Randall-Sundrum exponential warp-factor, we assume the following brane action:
        \begin{equation} 
        S_{brane}=-6\int d^5x\sqrt{-e_4}k(\delta(y)-\delta(y-\pi r_c)),
        \end{equation} 
        where we have defined $k=\frac{2\sqrt{2}}{3} g|R|$. Notice, that in this case we have broken supersymmetry on the second brane 
\cite{Gherghetta:2002nr},\cite{Brax:2001xf}.

        Let us investigate the spectrum of the effective theory. Consider small fluctuations of the 4d components of the metric tensor around RS vacuum solution: $g_{\mu\nu}(x^{\rho},y)=e^{-2k|y|}\eta_{\mu\nu}+\phi_{h}(y)h_{\mu\nu}(x^{\rho})$, where $h_{\mu\nu}(x^{\rho})$ is a 4d wave function ($\partial^{\rho}\partial_{\rho} h_{\mu\nu}=m^{2}h_{\mu\nu}$) in the gauge $\partial^{\mu}h_{\mu\nu}=h_{\mu}^{\mu}=0$. Linearized Einstein equations reduce to:
        \begin{equation} \label{kkflatgrawiton}
        \frac{1}{2}\phi_{h}''-2k^{2}\phi_{h}+2k(\delta(y)-\delta(y-\pi r_{c}))\phi_{h}+\frac{1}{2}e^{2k|y|}m^{2}\phi_{h}=0.
        \end{equation}   
        Massless and massive solution are $\phi_h=A_{0}e^{-2k|y|}$ and $\phi_h=A_{m}J_{2}(\frac{m}{k}e^{k|y|})+B_{m}Y_{2}(\frac{m}{k}e^{k|y|})$ respectively. $J_{2}$ and $Y_{2}$ are Bessel and Newman function of the second kind. Matching delta function at the fixed points determines quantization mass condition:
        \begin{equation} \label{masygrawitonuflat}
        J_{1}(\frac{m}{k})Y_{1}(\frac{m}{k}e^{k\pi r_{c}})-Y_{1}(\frac{m}{k})J_{1}(\frac{m}{k}e^{k\pi r_{c}})=0 \ .
        \end{equation} 
        
\begin{figure}[h]
        \begin{center}
        \epsfig{file=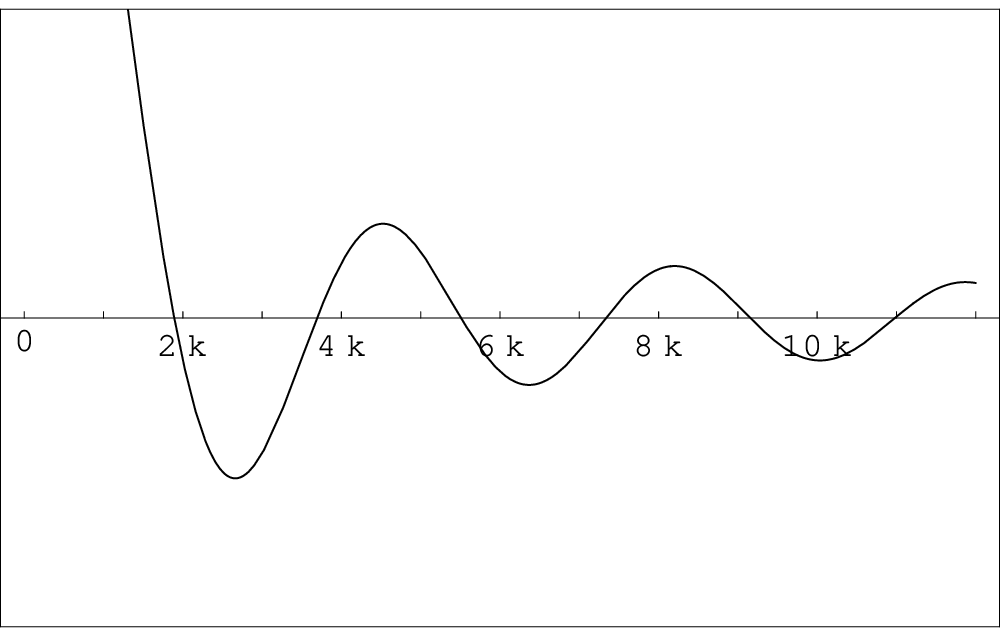, width=.5 \linewidth}
        \caption{LHS of the equation (\ref{masygrawitonuflat}) as a function of the mass parameter m. Zeros denote mass spectrum of the graviton ($k\pi r_{c}=1$). \label{masygrav}}  
        \end{center}
        \begin{center}                                        
        \epsfig{file=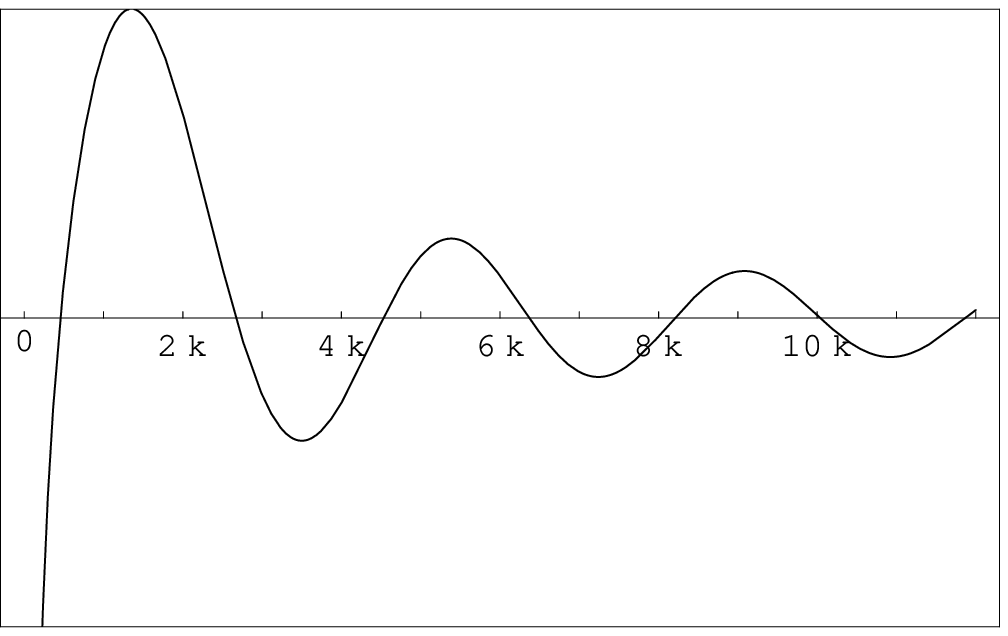, width=.5 \linewidth}             
        \caption{LHS of the equation (\ref{masygrawitinaflat}) as a function of the mass parameter m. Zeros denote mass spectrum of the gravitino ($k\pi r_{c}=1$). \label{masygravit}}
        \end{center}
        \end{figure}

        To compare KK masses in bosonic and fermionic sectors, let us focus 
on the gravitino equation of motion:
        \begin{equation} 
        \gamma^{\alpha\beta\gamma}D_{\beta}\Psi_{\gamma}^{A}+{\rm i}\sqrt{2}gP^{A}_{B}\gamma^{\alpha\gamma}\Psi_{\gamma}^{B}=0,
        \end{equation} 
        where $D_{\beta}\Psi_{\gamma}^{A}=(\partial_{\beta}+\frac{1}{4}(\omega_{\beta})_{mn}\gamma^{mn})\Psi_{\gamma}^{A}$ are covariant derivative and $(\omega_{\beta})_{mn}$ is a spinor connection. In the flat background and in gauge $\Psi_{5}=0$ we can write:
        \begin{equation} 
        \gamma^{\mu\rho\nu}\partial_{\rho}\Psi_{\nu}^{A}-\gamma^{\mu\nu}\gamma^{5}\partial_{5}\Psi_{\nu}^{A}+k\epsilon\gamma^{\mu\nu}\gamma^{\hat{5}}\Psi_{\nu}^{A}-\sqrt{2}gR\epsilon(\sigma_3)^{A}_{B}\gamma^{\mu\nu}\Psi_{\nu}^{B}=0,
        \end{equation} 
        where a hat denotes a flat space Dirac matrix. In the next step we factorize the gravitino wave function as follows:
        \begin{eqnarray}
        &(\Psi^{1}_{\mu})_R=\phi_{\psi}^{+}(y)(\psi_{\mu}^{+})_{R}(x^{\rho})\ ,\quad(\Psi^{1}_{\mu})_L=\phi_{\psi}^{-}(y)(\psi_{\mu}^{+})_{L}(x^{\rho})\ ,&\nonumber\\&(\Psi^{2}_{\mu})_L=-\phi_{\psi}^{+}(y)(\psi_{\mu}^{-})_{L}(x^{\rho})\ ,\quad(\Psi^{2}_{\mu})_R=\phi_{\psi}^{-}(y)(\psi_{\mu}^{-})_{R}(x^{\rho})\ ,&
        \end{eqnarray}
        where 4d gravitini satisfy the Rarita-Schwinger equations with the mass parameter $m$
        \begin{eqnarray}
        &\gamma^{\mu\nu\rho}\partial_{\nu}\psi^{+}_{\rho}=m\gamma^{\mu\rho}\psi^{+}_{\rho}&\nonumber\\&\gamma^{\mu\nu\rho}\partial_{\nu}\psi^{-}_{\rho}=-m\gamma^{\mu\rho}\psi^{-}_{\rho}&\ .
        \end{eqnarray}
        Gravitino equation of motion yields
        \begin{eqnarray} \label{equationgravitflat}
        &\phi^{+\prime}_{\psi}+\frac{1}{2}k\epsilon\phi^{+}_{\psi}-me^{k|y|}\phi^{-}_{\psi}=0&\nonumber\\&\phi^{-\prime}_{\psi}-\frac{5}{2}k\epsilon\phi^{-}_{\psi}+me^{k|y|}\phi^{+}_{\psi}=0&\ .
        \end{eqnarray}
        For $m=0$ the solutions are $\phi_{\psi}^{+}=A^{+}_{0}\epsilon_{\pi}e^{-\frac{k}{2}|y|}$ and $\phi_{\psi}^{-}=A^{-}_{0}\epsilon_{0}e^{\frac{5k}{2}|y|}$, where the `flipped' step functions $\epsilon_{0}(y)=\epsilon(\frac{y}{2})$ and $\epsilon_{\pi}(y)=\epsilon(\frac{y+\pi r_c}{2})$ are needed to satisfy conditions (\ref{plaskiz2}). In this case both $\partial_{5}\phi_{\psi}^{+}$ and $\partial_{5}\phi_{\psi}^{-}$ contain delta functions at points $y=\pi r_{c}$ and $y=0$ respectively, and those cannot be cancelled in the equation of motion (\ref{equationgravitflat}). Hence, the massless modes do not exist in the model.
Solutions for nonzero modes are
        \begin{eqnarray}
        &\phi_{\psi}^{+}=e^{\frac{3}{2}k|y|}\epsilon_{\pi}\left(A_{m}J_{2}(\frac{m}{k}e^{k|y|})+B_{m}Y_{2}(\frac{m}{k}e^{k|y|})\right)&\nonumber\\&\phi_{\psi}^{-}=e^{\frac{3}{2}k|y|}\epsilon_{0}\left(A_{m}J_{1}(\frac{m}{k}e^{k|y|})+B_{m}Y_{1}(\frac{m}{k}e^{k|y|})\right)&\ .
        \end{eqnarray}
        Matching conditions imply vanishing of the functions $\phi_{\psi}^{+}$ and $\phi_{\psi}^{-}$ at the points $y=\pi r_{c}$ and $y=0$ respectively. Associated mass quantization condition reads:
        \begin{equation} \label{masygrawitinaflat}
        J_{1}(\frac{m}{k})Y_{2}(\frac{m}{k}e^{k\pi r_{c}})-Y_{1}(\frac{m}{k})J_{2}(\frac{m}{k}e^{k\pi r_{c}})=0\ .
        \end{equation} 
The quantization of the mass parameters for gravitons and gravitini can be 
read from the figures \ref{masygrav} and \ref{masygravit}. The spectra, together with the absence of the zero modes for gravitini imply broken supersymmetry
(the background is flat here).  
        
\subsection{$AdS_4$ compactification of the pure supergravity with flipped boundary conditions (super-bigravity)}

In section \ref{pomarolr} we saw, that in the Randall-Sundrum model with flipped boundary conditions, supersymmetry is broken in the effective 4d theory.
 The reason for this fact is twofold: flipped boundary conditions and 
explicitly broken 
supersymmetry at the point $y=\pi r_c$.  To procede let us go on to
 the locally  supersymmetric model with a flip along the fifth dimension. 
The price for local supersymmetry and the trouble one encounters is 
the nonzero curvature in 4d sections. 
Let us take the supergravity action with the prepotential of the form: $P=g_{R}\epsilon(y){\rm i}\sigma_3R+g_{S}{\rm i}\sigma_1S$, and the brane action  required by supersymmetry:
        \begin{equation} 
        S_{brane}=-6\int d^5x\sqrt{-e_4}kT(\delta(y)+\delta(y-\pi r_c)),
        \end{equation}   
        where 
$       k=\frac{2\sqrt{2}}{3}\sqrt{g_{R}^{2}R^{2}+g_{S}^{2}S^{2}},$ and 
$       T=g_{R}|R|/\sqrt{g_{R}^{2}R^{2}+g_{S}^{2}S^{2}}.$ One should 
notice that brane tensions have the same sign. As a consequence gravitational background has no flat 4d Minkowski foliation, and the consistent 
solution is that of $AdS_4$ branes:
        \begin{equation} 
        ds^{2}=a^{2}(y)\bar{g}_{\mu\nu}dx^{\mu}dx^{\nu}+dy^{2}\ ,
        \end{equation}  
        where 
        \begin{equation} 
        a(y)=\frac{\sqrt{-\bar{\Lambda}}}{k}\cosh\left(k|y|-\frac{k\pi r_{c}}{2}\right)\ ,
        \end{equation} 
        and $\bar{g}_{\mu\nu}dx^{\mu}dx^{\nu}=\exp(-2\sqrt{-\bar{\Lambda}}x_{3})(-dt^{2}+dx^{2}_{1}+dx_{2}^{2})+dx_{3}^{2}$ is the four dimensional $AdS$ metric. 

The radius of the fifth dimension is determined by brane tensions:
        \begin{equation} 
        k\pi r_{c}=\ln\left(\frac{1+T}{1-T}\right)\ .
        \end{equation} 
        Normalization $a(0)=1$ leads to the fine tuning relation $\bar{\Lambda}=(T^{2}-1)k^{2}<0$. As in the previous paragraph we look at small fluctuations around vacuum metric: $g_{\mu\nu}(x^{\rho},y)=a^{2}(y)\bar{g}_{\mu\nu}+\phi_{h}(y)h_{\mu\nu}(x^{\rho})$, where $h_{\mu\nu}(x^{\rho})$ is a 4d wave function in $AdS_{4}$  background ($(\Box_{AdS}+ 2\bar{\Lambda})h_{\mu\nu}=m^{2}h_{\mu\nu}$ \cite{Alonso-Alberca:2000ne},\cite{Karch:2000ct}). The analog of the equation (\ref{kkflatgrawiton}) reads:
        \begin{equation} \label{kkadsgrawiton}
        \frac{1}{2}\phi_{h}''-2k^{2}\phi_{h}+2k\tanh\left(\frac{k\pi r_{c}}{2}\right)(\delta(y)+\delta(y-\pi r_{c}))\phi_{h}+\frac{1}{2}a^{2}(y)(m^{2}-2\bar{\Lambda})\phi_{h}=0.
        \end{equation}   
        It is easy to check, that the massless mode $\phi_h=A_{0}\cosh^{2}(k|y|-k\pi r_{c}/2)$ satisfies the equation of motion in the bulk and the 
boundary conditions. Massive modes can be written as    
        \begin{eqnarray}
        \phi_h=&A_{m}{\rm LP}\left(\frac{1}{2}\left(-1+\sqrt{1+4\bar{m}^{2}}\right),2,\tanh\left(k|y|-\frac{k\pi r_{c}}{2}\right)\right)+&\nonumber\\&+B_{m}{\rm LQ}\left(\frac{1}{2}\left(-1+\sqrt{1+4\bar{m}^{2}}\right),2,\tanh\left(k|y|-\frac{k\pi r_{c}}{2}\right)\right)&\ ,
        \end{eqnarray}  
        where ${\rm LP}(m,n,x)$ and ${\rm LQ}(m,n,x)$ are associated Legendre functions of the first and second kind respectively. We have introduced the new symbol $\bar{m}=\sqrt{-m^{2}/\bar{\Lambda}+2}$. Matching delta functions at fixed points leads to the  following mass quantization condition
        \begin{eqnarray} \label{warunkinamasyadsgrawiton}
        0=&\left(2t{\rm LQ}(,,-t) +c{\rm LQ}'(,,-t)\right)\left(-2t{\rm LP}(,,t) +c{\rm LP}'(,,t)\right)+&\nonumber\\&-\left(2t{\rm LP}(,,-t)+ c{\rm LP}'(,,-t)\right)\left(-2t{\rm LQ}(,,t)+c{\rm LQ}'(,,t)\right)&,
        \end{eqnarray}  
        where we have introduced notation $t=\tanh(k\pi r_{c}/2)$ and $c=\cosh^{-2}(k\pi r_{c}/2)$.

        The gravitino equation of motion in the AdS background reads
        \begin{equation} 
        \gamma^{\mu\rho\nu}\nabla_{\rho}\Psi_{\nu}^{A}-\gamma^{\mu\nu}\gamma^{5}\partial_{5}\Psi_{\nu}^{A}+k\epsilon\gamma^{\mu\nu}\gamma^{\hat{5}}\Psi_{\nu}^{A}-\sqrt{2}\left(g_{R}R\epsilon(\sigma_3)^{A}_{B}+g_{S}S(\sigma_1)^{A}_{B}\right)\gamma^{\mu\nu}\Psi_{\nu}^{B}=0\ ,
        \end{equation}
        where $\nabla_{\mu}$ denotes $AdS_4$ covariant derivative.
        Notice that the prepotential mixes $\Psi_{\mu}^{1}$ and $\Psi_{\mu}^{2}$ fields. To eliminate this mixing, let us define the following functions
        \begin{eqnarray}
        &\Psi^{+}_{\mu}=g_S|S|\Psi_{\mu}^1+\epsilon(\sqrt{g_R^2R^{2}+g_S^2S^{2}}-g_R|R|)\Psi_{\mu}^2\ ,&\nonumber\\&\Psi^{-}_{\mu}=g_S|S|\Psi_{\mu}^2-\epsilon(\sqrt{g_R^2R^{2}+g_S^2S^{2}}-g_R|R|)\Psi_{\mu}^1\ .&
        \end{eqnarray}
        The factorization
        \begin{eqnarray}
        &(\Psi^{+}_{\mu})_R=\phi_{\psi}^{+}(y)(\psi_{\mu}^{+})_{R}(x^{\rho})\ ,\quad(\Psi^{+}_{\mu})_L=\phi_{\psi}^{-}(y)(\psi_{\mu}^{+})_{L}(x^{\rho})\ ,&\nonumber\\&(\Psi^{-}_{\mu})_L=-\phi_{\psi}^{+}(y)(\psi_{\mu}^{-})_{L}(x^{\rho})\ ,\quad(\Psi^{-}_{\mu})_R=\phi_{\psi}^{-}(y)(\psi_{\mu}^{-})_{R}(x^{\rho})\ ,&
        \end{eqnarray}
        where 4d gravitini satisfy the Rarita-Schwinger equations in $AdS_4$:
        \begin{eqnarray}
        &\gamma^{\mu\nu\rho}\nabla_{\nu}\psi^{+}_{\rho}=(m-\sqrt{-\bar{\Lambda}})\gamma^{\mu\rho}\psi^{+}_{\rho},&\nonumber\\&\gamma^{\mu\nu\rho}\nabla_{\nu}\psi^{-}_{\rho}=-(m-\sqrt{-\bar{\Lambda}})\gamma^{\mu\rho}\psi^{-}_{\rho},&
        \end{eqnarray}
        leads to the equation
        \begin{eqnarray} \label{equationgravitads}
        &\phi^{+\prime}_{\psi}+k\epsilon\left(\tanh\left(k|y|-\frac{k\pi r_{c}}{2}\right)+\frac{3}{2}\right)\phi^{+}_{\psi}-a^{-1}(y)(m-\sqrt{-\bar{\Lambda}})\phi^{-}_{\psi}=0&\nonumber\\&\phi^{-\prime}_{\psi}+k\epsilon\left(\tanh\left(k|y|-\frac{k\pi r_{c}}{2}\right)-\frac{3}{2}\right)\phi^{-}_{\psi}+a^{-1}(y)(m-\sqrt{-\bar{\Lambda}})\phi^{+}_{\psi}=0&.
        \end{eqnarray}
The boundary conditions are imposed by the action of the        
$Z_2$ in the fermionic sector (\ref{plaskiz2}). One needs to demand that the fields 
$(\Psi_{\mu}^2)_R$ and $(\Psi_{\mu}^1)_R$ vanish at the points $y=0$ and $y=\pi r_{c}$ respectively. This implies
        \begin{equation} \label{wargravitinoads}
        \phi_{\psi}^{+}(0)=-e^{\frac{k\pi r_{c}}{2}}\phi_{\psi}^{-}(0)\ ,\qquad \phi_{\psi}^{-}(\pi r_{c})=e^{\frac{k\pi r_{c}}{2}}\phi_{\psi}^{+}(\pi r_{c})\ .
        \end{equation} 
This condition removes the mode $m=0$ from the spectrum. 
Solutions of the equation
 (\ref{equationgravitads}) with a nonzero mass can be written down as follows
        \begin{eqnarray} \label{solutionmassivegravitads}
        &\phi^{+}_{\psi}=e^{-\frac{k}{2}\left(|y|-\frac{\pi r_c}{2}\right)}\left(A_m(1-z){\;}_{2}F_{1}[-M, M, 2,\frac{1-z}{2}]+B_m\frac{M}{4}(1+z)^{2}{\;}_2F_{1}[1-M,1+M,3,\frac{1+z}{2}]\right)&\nonumber\\&\phi^{-}_{\psi}=e^{\frac{k}{2}\left(|y|-\frac{\pi r_c}{2}\right)}\left(A_m\frac{M}{4}(1-z)^{2}{\;}_2F_{1}[1-M,1+M,3,\frac{1-z}{2}]+B_m(1+z){\;}_2F_{1}[-M, M, 2,\frac{1+z}{2}]\right),&\nonumber\\&&\ 
        \end{eqnarray} where we have introduced $z=\tanh\left(k|y|-\frac{k\pi r_c}{2}\right)$ and $M=(m/\sqrt{-\bar{\Lambda}})-1$. The symbol ${\;}_{2}F_{1}[a, b, c,x]$ denotes a hypergeometric function.
The condition (\ref{wargravitinoads}) takes on the following form
        \begin{eqnarray} \label{wargravitinoads2}
        &\left(\frac{M}{4}(1+t)^{2}F[3,\frac{1+t}{2}]+(1+t)F[2,\frac{1+t}{2}]\right)\left((1+t)F[2,\frac{1+t}{2}]-\frac{M}{4}(1+t)^{2}F[3,\frac{1+t}{2}]\right) +&\nonumber\\&+\left((1-t)F[2,\frac{1-t}{2}]-\frac{M}{4}(1-t)^{2}F[3,\frac{1-t}{2}]\right)\left((1-t)F[2,\frac{1-t}{2}]+\frac{M}{4}(1-t)^{2}F[3,\frac{1-t}{2}]\right)=&\nonumber\\&=0,&
        \end{eqnarray}
        where for simplicity we have introduced the notation  $F[2,\frac{1\pm t}{2}]={\;}_{2}F_{1}[-M, M, 2,\frac{1\pm t}{2}]$ and $F[3,\frac{1\pm t}{2}]={\;}_{2}F_{1}[1-M,1+M, 3,\frac{1\pm t}{2}]$. 
        \begin{figure}[h]
        \begin{center}
        \epsfig{file=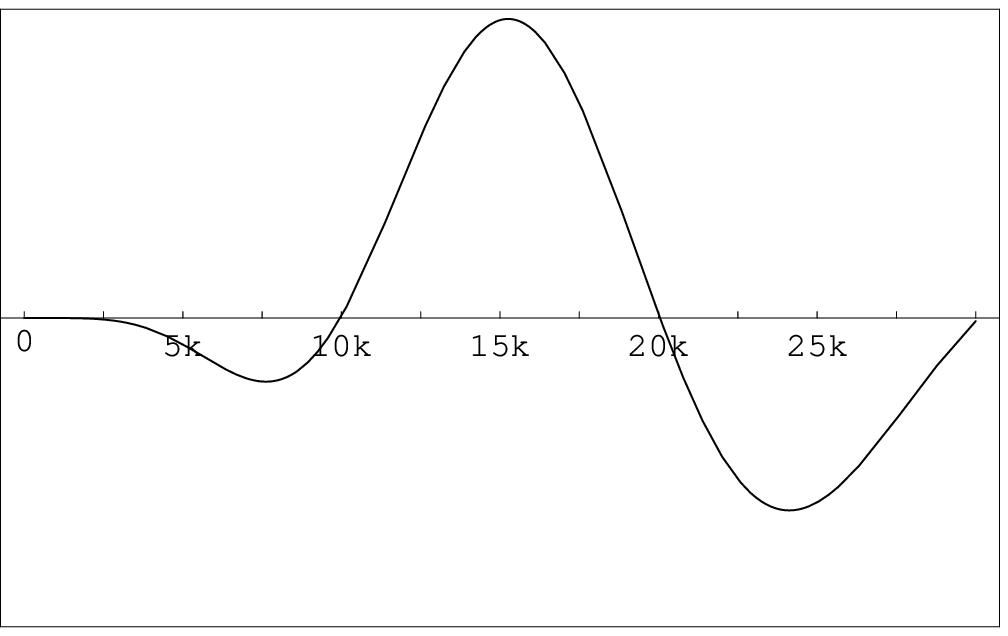, width=.45 \linewidth}
        \caption{The zeros corresponds to the mass spectrum of the graviton, the quantization condition is plotted as the function of the variable $m\cosh(\pi r_c/2)$ 
($k r_{c}=0.1$). \label{masygravitonu0.1}}  
        \end{center}
        \begin{center}                                        
        \epsfig{file=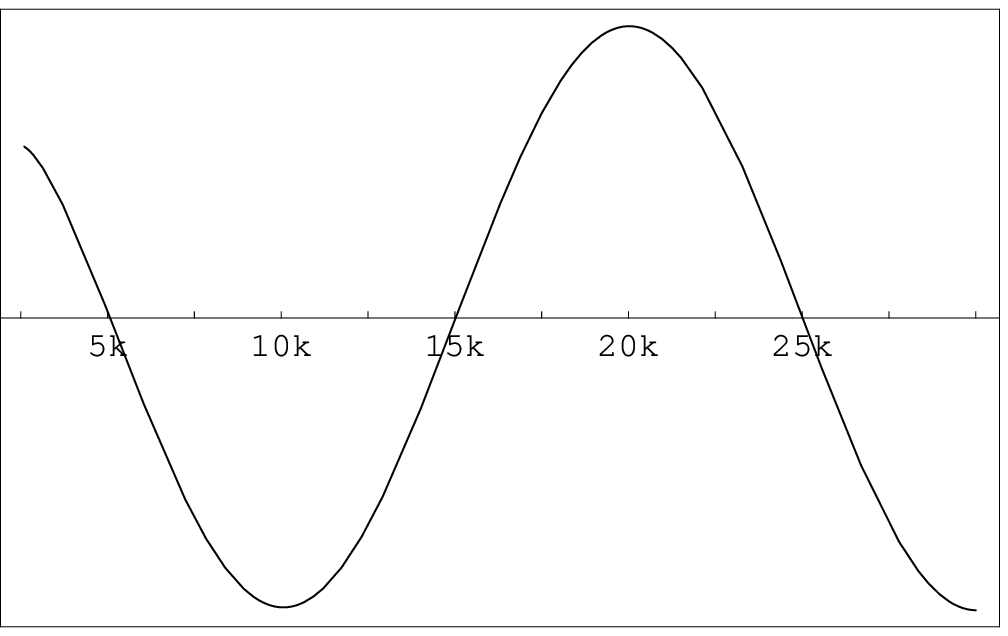, width=.45 \linewidth}             
        \caption{The zeros correspond to the mass spectrum of the gravitino (condition(\ref{wargravitinoads2})),  the quantization condition is plotted as the function of the variable $M$ ($kr_{c}=0.1$). \label{masygravitina0.1}}
        \end{center}    
        \end{figure}       
In the figures we have shown the mass quantization that follows from the 
conditions  (\ref{warunkinamasyadsgrawiton}) and (\ref{wargravitinoads2}). 
The spectrum of the gravitino mass parameter $m$ is shifted by approximately half a distance between consecutive zeros with respect to the mass spectrum of the graviton. 
It turns out that one can compute analytically  the graviton and gravitini mass spectra in the limiting cases of a large extra dimension ($kr_{c} \gg 1$) and in the case of a small
extra dimension ($k r_c \ll 1$) (see Appendix). In the regime $kr_{c} \gg 1$ we obtain the ultra-light graviton mode
        \begin{equation} 
	m^{2}_{light}\approx  12k^{2} e^{-k\pi r_{c}}\cosh^{-2}(k\pi r_{c}/2)\ ,
	\end{equation} 
and heavy modes
        \begin{equation} \label{lah}
	  m^{2}_h \approx k^2 (-2+n+n^{2}) \cosh^{-2}(k\pi r_{c}/2)=(-2+n+n^{2})|\bar{\Lambda}|\ ,
	\end{equation}
for $n>1$. For gravitini we obtain: 
\begin{equation} \label{lag}
m^{2}_f \approx k^2  (n+1)^2 \cosh^{-2} (k\pi r_{c}/2)= (n+1)^2 |\bar{\Lambda}|\ .
\end{equation}
For large $n$ we can write for gravitons: $m\approx k(1/2+n)\cosh^{-1}(k\pi r_{c}/2)$. 

In the limit $k r_c \ll 1$, the equations (\ref{kkadsgrawiton}) and (\ref{equationgravitads}) together 
with the assumptions $kr_c \ll 1$, $\bar{m}\approx\sqrt{m^2/k^2+2}\gg 1$ and $M\approx m/k \gg 1$ give the following mass quantization
\begin{equation} \label{smah}
m^{2}_h=\frac{n^2}{r^{2}_c},  
\end{equation}
and 
\begin{equation} \label{smag}
m^{2}_{\psi}=\frac{1}{r^{2}_c}(\frac{1}{2}+n)^{2}.
\end{equation}
The approximate spectra for the gravitini masses that we have just obtained 
can be compared to the spectra of the massive spin-2 states belonging to the 
$AdS_4$
supermultiplets discussed earlier given the $AdS_4$ mass formula 
$m^2 = C_2 (E_0,s) - C_2(s+1,s)= E_0 (E_0 -3) -(s+1)(s-2)$ for representations 
$D(E_0, s)$. In the limit of dimensional reduction \cite{Karch:2000ct} 
this implies the 
spin-2 and spin-3/2 spectra $m^{2}_{2,n} = (E_0 + 1/2 +n)
(E_0 -5/2 +n)$ and $m^{2}_{3/2,n}= (E_0 +n)(E_0 + n -3) +5/4$, 
${m'}^{2}_{3/2,n}= (E_0 +n+1)(E_0 + n -2) +5/4$, 
for some $E_0$ and $n=0,1,2,...$ (in units of $\sqrt{-\bar{\Lambda}}$). 
The above mass formula fits the limiting ($k r_c \gg 1$) spectra of graviton 
(except the first massive mode) and gravitino masses 
(\ref{lah}) and (\ref{lag})
if $E_0 = 3/2$, but this value does not correspond to a unitary supermultiplet, since the necessary condition $E_0 > s+1$ \cite{Heidenreich:rz} 
is not fulfilled for $s=3/2$ and $E_0=3/2$. 
The natural value for dimensional reduction $5d \rightarrow 4d$ would be $E_0 =3$. This gives 
$m^{2}_{2,n}=n^2 + 4n + 7/4$, $m^{2}_{3/2,n}=(n+3/2)^2 -1$ and ${m'}^{2}_{3/2,n} = (n+5/2)^2 -5$,
again in clear mismatch with (\ref{lah}) and (\ref{lag}).
It is also clear that the graviton mass spectrum 
for a finite $k r_c$ differs from the supersymmetric one.  
In the case where the $r_c$ is much smaller than the curvature radius,
the spectrum of gravitons and gravitini approaches the usual, flat space, 
KK form with gravitini masses shifted with respect to these of the gravitons. 
Also in this limit the spectrum is clearly nonsupersymmetric, and the shift 
is due solely to the twisted boundary conditions.   

\begin{figure}[h]
        \begin{center}
        \epsfig{file=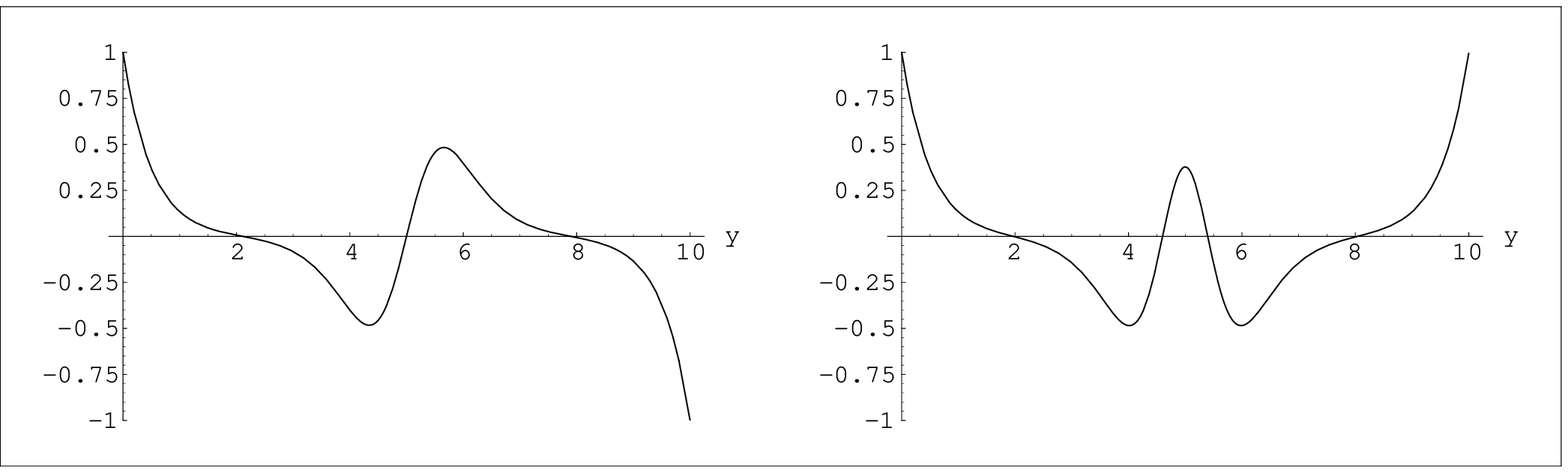, width=.8 \linewidth}
        \caption{The third and the fourth graviton modes ($k\pi r_{c}=10$, $m=3.16k$ and $m=4.24k$). \label{modygravito}}  
        \end{center}  
%	\end{figure}   
%	\begin{figure}[h]
        \begin{center}                                        
        \epsfig{file=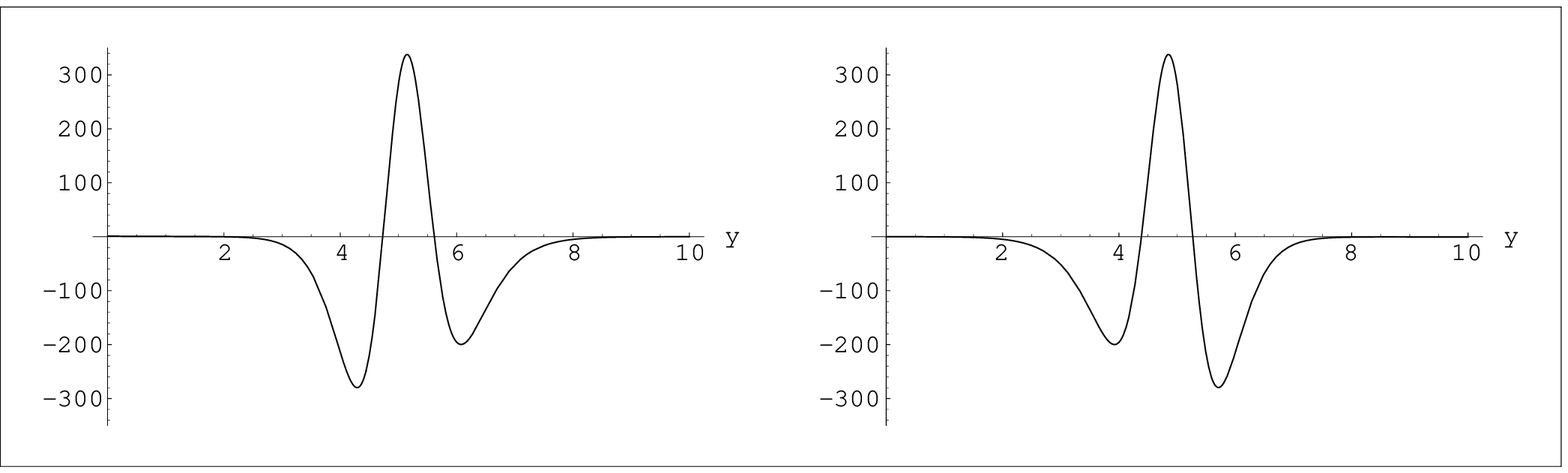 , width=.8 \linewidth}          
        \caption{The fourth modes of the gravitini $\phi_{\psi}^+$ and $\phi_{\psi}^-$ ($k\pi r_{c}=10$, $m=4k$). \label{modygravit1}}
        \end{center} 
%	\begin{center}                                        
%       \epsfig{file=gravitino2_mn_10_4.ps, width=.4 \linewidth}             
%      \caption{Fourth gravitini mode ($k\pi r_{c}=10$, $m=4k$) \label{modygravit2}}
%     \end{center}
        \end{figure}   

One can see that even in the limit $r_c \gg 1/k$ supersymmetry is not restored,
and the branes do not decouple like in the supersymmetric Randall-Sundrum 
case. The nondecoupling may also be seen from the shape of the wave 
functions of the massive modes (figure \ref{modygravito} and \ref{modygravit1}). 

\begin{figure}[h] 
        \begin{center}
        \epsfig{file=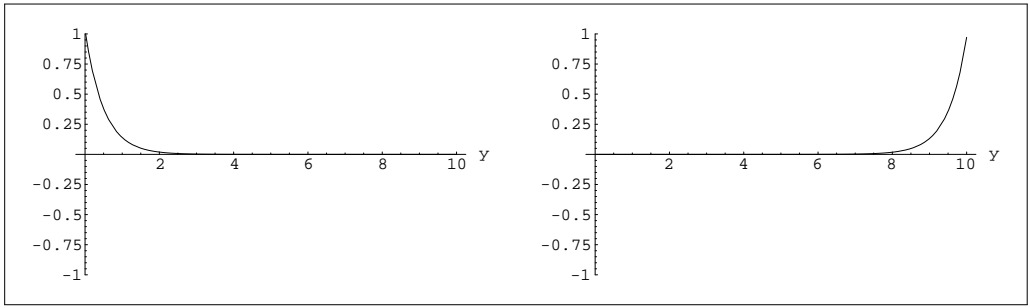, width=.8 \linewidth} 
	  \caption{``Left'' and ``right'' light graviton modes ($k\pi r_{c}=10$). \label{gravleft}}  
        \end{center}
	\end{figure}

	However, when $kr_c$ goes to infinity, $m_{light}\rightarrow 0$. In 
such a regime we can take the following linear combinations of the ultra-light mode and the zero modes:
	\begin{equation}
	\phi_{left}(y)=\frac{1}{2}\left(\frac{\phi_0(y)}{\phi_0(0)}+\frac{\phi_{light}(y)}{\phi_{light}(0)}\right)\ ,\quad\phi_{right}(y)=\frac{1}{2}\left(\frac{\phi_0(y)}{\phi_0(0)}-\frac{\phi_{light}(y)}{\phi_{light}(0)}\right)\ .
	\end{equation} 
	Then $\phi_{left}$ ($\phi_{right}$) is localized on the brane at $y=0$ ($y=\pi r_c$) and vanishes on the brane located at $y=\pi r_c$ ($y=0$). Hence, effectively we have only one zero mode on each brane (figure \ref{gravleft}).  

To summarize the discussion of the supersymmetry breakdown in the case of the flipped supergravity let us inspect the equation for the Killing spinors: 
          \begin{eqnarray} 
	  &\left(\frac{a'}{a}+\frac{2\sqrt{2}}{3}g_{1}\epsilon(y)|R|\right)\epsilon_+^A+\frac{2\sqrt{2}}{3}\gamma_5g_{2}|S|(\sigma^1)^A_B\epsilon_-^B=0&\nonumber\\ &\left(\frac{a'}{a}-\frac{2\sqrt{2}}{3}g_{1}\epsilon(y)|R|\right)\epsilon_-^A+\frac{2\sqrt{2}}{3}\gamma_5g_{2}|S|(\sigma^1)^A_B\epsilon_+^B=0&\ ,
	  \end{eqnarray}
where $\epsilon^{A}_{\pm} = 1/2 (\delta^{A}_{B} \pm \gamma_5 Q^{A}_{B}) \epsilon^{B}$. These 
equations result in the condition
	  \begin{equation} 
	  \left(\left(\frac{a^{\prime 2}}{a^{2}} - \frac{8}{9}g_{1}^{2}R^{2}\right)-\frac{8}{9}g_{2}^{2}S^{2}\right)\epsilon_{\pm}^A=0\ ,
	  \end{equation} 
and together with Einstein equations this implies that for non-vanishing $S$ there are no nontrivial solutions of the Killing equation.
\section{Summary and conclusions}
We have shown that flipped and gauged five-dimensional supergravity 
is closely related to the Scherk-Schwarz mechanism of symmetry breakdown. 
In this case the Scherk-Schwarz redefinition of fields connects two phases of the model. 
One phase is such that supersymmetry is broken spontaneously, in the sense that there do not 
exist vacua preserving some of the supercharges. In fact, one cannot undo this breakdown in a continous way, since the choice of the projectors on both branes is a discrete one - one cannot 
deform continously $Q$ into $-Q$ within the model. In particular, in the limit $r_c \rightarrow 0$ 
all gravitini (and all supercharges) get projected away. In the second, Scherk-Schwarz phase,
linear supersymmetry is not realized explicitly in the Lagrangian, hence one finds susy breaking masses and potential terms in the bulk and/or on the branes.  
However, the physics of the two phases has to be the same, as they are 
related by a mere redefinition of variables.    

We have found that the simple flipped 5d supergravity is a supersymmetrization of the 
$(++)$ bigravity with two positive tension branes. In the limit of the large interbrane separation 
there exists a ultra-light massive graviton mode in addition to the exactly massless mode
(but there is no a nearly degenerate superpartner). 

As an example the Scherk-Schwarz terms for gauged supergravity coupled to bulk matter have been worked out. In addition it has been shown that the five-dimensional Horava-Witten model is not of the 
Scherk-Schwarz type, since flipping of supersymmetry requires `wrong' boundary terms on the branes.

In the class of models discussed here it is the $AdS_4$ background that appears naturally as a static solution of the equations of motion. However, firstly, there exist nearby time-dependent solutions 
leading to Robertson-Walker type cosmology on branes, and secondly, in more realistic models the gravitational background we have described shall be further perturbed by nontrivial gauge and matter 
sectors living on the branes.

\vskip 1cm

\noindent{\large \bf Acknowledgments}

\vskip .3cm

\noindent
The authors thank P. Brax and S. Theisen for very helpful discussions. Z. L. thanks CERN Theory Division where part of this project was done.  
This work  was partially supported  by the EC Contract
HPRN-CT-2000-00152 for years 2000-2004, by the Polish State Committee for Scientific Research grant KBN 5 P03B 119 20 for years 2001-2002, and by 
POLONIUM 2002.
 
\vspace{1cm}
\newpage
\appendix
\noindent{\large \bf Appendix }
\setcounter{equation}{0}
\renewcommand{\theequation}{A.\arabic{equation}}
\vspace{0.3cm}
%\section{Analytic solutions}

In the regime $kr_{c} \gg 1$ we have used the following approximations:
        \begin{eqnarray} \label{aproxlarger}
        &&{\rm LP}(,,-t)\approx \frac{1}{\pi}\cos\left(\frac{1}{2}\sqrt{1+4\bar{m}^{2}}\pi\right)\left(e^{k\pi r_{c}}+\bar{m}^{2}-1\right)+O(e^{-k\pi r_{c}})\ ,\nonumber\\
        &&c{\rm LP}'(,,-t)\approx -\frac{2}{\pi}\cos\left(\frac{1}{2}\sqrt{1+4\bar{m}^{2}}\pi\right)e^{k\pi r_{c}}+O(e^{-k\pi r_{c}})\ ,\nonumber\\
        &&{\rm LP}(,,t)\approx \frac{1}{2} \bar{m}^{2}(\bar{m}^{2}-2)e^{-k\pi r_{c}}+O(e^{-2k\pi r_{c}})\ ,\nonumber\\
        &&c{\rm LP}'(,,t)\approx - \bar{m}^{2}(\bar{m}^{2}-2)e^{-k\pi r_{c}}+O(e^{-2k\pi r_{c}})\ ,\nonumber\\
        &&{\rm LQ}(,,-t)\approx -\frac{1}{2}\sin\left(\frac{1}{2}\sqrt{1+4\bar{m}^{2}}\pi\right)\left(e^{k\pi r_{c}}+\bar{m}^{2}-1\right)+O(e^{-k\pi r_{c}})\ ,\nonumber\\
        &&c{\rm LQ}'(,,-t)\approx \sin\left(\frac{1}{2}\sqrt{1+4\bar{m}^{2}}\pi\right)e^{k\pi r_{c}}+O(e^{-k\pi r_{c}})\ ,\nonumber\\
        &&{\rm LQ}(,,t)\approx \frac{1}{2} \left(e^{k\pi r_{c}}+\bar{m}^{2}-1\right)+O(e^{-k\pi r_{c}})\ ,\nonumber\\
        &&c{\rm LQ}'(,,t)\approx e^{k\pi r_{c}}+O(e^{-k\pi r_{c}})\ ,\nonumber\\
        &&t\approx 1-2e^{-k\pi r_{c}}+O(e^{-2k\pi r_{c}})\ ,
        \end{eqnarray}
        and
        \begin{eqnarray} \label{aproxlarger2}
        &&(1+t)F[2,\frac{1+t}{2}]\approx -\frac{2}{(M^{2}-1)M\pi}\sin(M\pi)+O(e^{-k\pi r_{c}})\ ,\nonumber\\
        &&\frac{M}{4}(1+t)^{2}F[3,\frac{1+t}{2}]\approx -\frac{2}{(M^{2}-1)\pi}\sin(M\pi)+O(e^{-k\pi r_{c}})\ ,\nonumber\\
        &&(1-t)F[2,\frac{1-t}{2}]\approx 2e^{-k\pi r_{c}}+O(e^{-2k\pi r_{c}})\ ,\nonumber\\
        &&\frac{M}{4}(1-t)^{2}F[3,\frac{1-t}{2}]\approx Me^{-2k\pi r_{c}}+O(e^{-3k\pi r_{c}})\ .
        \end{eqnarray}
Then the conditions (\ref{warunkinamasyadsgrawiton}) and (\ref{wargravitinoads2}) reduce down to
        \begin{equation} 
        \cot\left(\frac{1}{2}\sqrt{9+4m^{2}}\pi\right)=\pi m^{2}\frac{m^2+2}{1-m^2}e^{-k\pi r_{c}}\approx 0\ ,
        \end{equation} 
        and
        \begin{equation} 
        \sin^{2}(M\pi)=\pi^{2} M^{2}(M^{2}-1)e^{-2k\pi r_{c}}\approx 0
        \end{equation} 
        respectively. Hence, we obtain the  mass quantization for gravitons: 
\begin{equation} \label{lahap}
m^{2}_h \approx k^2 (-2+n+n^{2}) \cosh^{-2}(k\pi r_{c}/2),
\end{equation}
 and for gravitini: 
\begin{equation} \label{lagap}
m^{2}_f \approx k^2  (n+1)^2 \cosh^{-2} (k\pi r_{c}/2). 
\end{equation}

%       \paragraph{Analytic solution for $kr_c<<1$.}
One can also obtain the analytic form of the spectrum in the limit $k r_c \ll 1$. In this regime 
%we expect masses much larger than $1$.
 the equations (\ref{kkadsgrawiton}) and (\ref{equationgravitads}) together 
with the assumptions $kr_c \ll 1$, $\bar{m}\approx\sqrt{m^2/k^2+2}\gg 1$ and $M\approx m/k \gg 1$ give
        \begin{equation}
          \phi''_h+(m^2-2 k^2)\phi_h=0\ ,
        \end{equation}
        and
        \begin{eqnarray}
          &\phi^{+\prime}_{\psi}+\frac{3}{2}k\epsilon\phi^{+}_{\psi}-m\phi^{-}_{\psi}=0&\ ,\nonumber\\&\phi^{-\prime}_{\psi}-\frac{3}{2}k\epsilon\phi^{-}_{\psi}+m\phi^{+}_{\psi}=0&
        \end{eqnarray}
        respectively. The solutions are very simple and take the form
        \begin{equation}
          \phi_h=A_m\cos\left[\sqrt(m^2-2k^2) \left(|y|-\frac{\pi r_c}{2}\right)\right ]+B_m\sin\left[\sqrt(m^2-2k^2) \left(|y|-\frac{\pi r_c}{2}\right)\right]\ ,
        \end{equation}
        for gravitons, and
        \begin{eqnarray}
          &&\phi^{+}_{\psi}=A_m\cos\left[\mu\left(|y|-\frac{\pi r_c}{2}\right)\right]+B_m\sin\left[\mu\left(|y|-\frac{\pi r_c}{2}\right)\right]\ ,\nonumber\\&&\phi^{-}_{\psi}=\left(\frac{3k}{2m}A_m+\frac{\mu}{m}B_m\right)\cos\left[\mu\left(|y|-\frac{\pi r_c}{2}\right)\right]+\left(\frac{3k}{2m}B_m-\frac{\mu}{m}A_m\right)\sin\left[\mu\left(|y|-\frac{\pi r_c}{2}\right)\right]\ ,\nonumber\\&&
        \end{eqnarray}
for gravitini. We have introduced above the notation $\mu=\sqrt{m^2-\frac{9}{4}k^2}$. Boundary conditions lead to the following mass quantizations
\begin{equation} \label{smahap}
m^{2}_h=\frac{n^2}{r^{2}_c},  
\end{equation}
and 
\begin{equation} \label{smagap}
m^{2}_{\psi}=\frac{1}{r^{2}_c}(\frac{1}{2}+n)^{2}.
\end{equation}

\end{document}